\documentclass[showpacs,preprintnumbers,amsmath,amssymb]{revtex4}
\begin{document}

\title{Stochastic Gravity: Beyond Semiclassical Gravity}

\author{E. Verdaguer}
\email{verdague@ffn.ub.es}

\affiliation{Departament de F\'{\i}sica Fonamental and CER en
Astrof\'\i sica, F\'\i sica de Part\'\i cules i Cosmologia,
Universitat de Barcelona, Av.~Diagonal 647, 08028 Barcelona,
Spain}


\begin{abstract}
The back-reaction of a classical gravitational field interacting
with quantum matter fields is described by the semiclassical
Einstein equation, which has the expectation value of the quantum
matter fields stress tensor as a source. The semiclassical theory
may be obtained from the quantum field theory of gravity
interacting with N matter fields in the large N limit. This theory
breaks down when the fields quantum fluctuations are important.
Stochastic gravity goes beyond the semiclassical limit and allows
for a systematic and self-consistent description of the metric
fluctuations induced by these quantum fluctuations. The
correlation functions of the metric fluctuations obtained in
stochastic gravity reproduce the correlation functions in the
quantum theory to leading order in an 1/N expansion. Two main
applications of stochastic gravity are discussed. The first, in
cosmology, to obtain the spectrum of primordial metric
perturbations induced by the inflaton fluctuations, even beyond
the linear approximation. The second, in black hole physics, to
study the fluctuations of the horizon of an evaporating black
hole.
\end{abstract}


\maketitle

%%%%%%%%%%%%%%%%%%%%%%

\section{Introduction}
\label{sec1}

There are two aspects to semiclassical gravity. On the one hand we
have quantum field theory in a curved spacetime which is now a
well understood and well defined theory both for free fields
\cite{Birrell 84} and interacting fields \cite{Wal06}. In this
theory the gravitational field is the classical field of general
relativity, the metric of the spacetime, and the quantum fields
are test fields which propagate in such a spacetime. Since the
spacetime is now dynamical it is not always possible to define a
physically meaningful vacuum state for the quantum field and when
this is possible in some ``initial" times it is usually unstable,
in the sense that it may differ from the vacuum state at latter
times, and spontaneous creation of particles occurs. Applications
of this in cosmology, such as particle production in expanding
Friedmann-Robertson-Walker models \cite{Parker,Par69}, and black hole
physics, such as Hawking radiation \cite{Hawking,Haw75}, are well known.
This is one aspect of the interaction of gravity with quantum
matter fields

The second aspect of this interaction is the back-reaction of the
quantum fields on the spacetime. Since the gravitational field
couples to the stress tensor of matter fields, the key object here
is the expectation value in a given quantum state of the stress
tensor of the quantum field, which is a classical observable.
However, since this object is quadratic in the field operator,
which is only well defined as a distribution on spacetime, it
involves ill defined quantities which translate into ultraviolet
divergences. To be able to define a physically meaningful
quantity a regularization and a renormalization procedure is
required. The ultraviolet divergences associated to the
expectation value of the stress tensor are also present in
Minkowski spacetime, but in a curved background the
renormalization procedure is more sophisticated since it needs to
preserve general covariance. A regularization procedure which is
specially adapted to the curved background is the so called
point-splitting method \cite{Christensen,Chr76,Chr78}.
The final expectation
value of the stress tensor using point splitting or any other
reasonable regularization technique such as dimensional
regularization, is essentially unique, modulo some terms which
depend on the spacetime curvature and which are independent of the
quantum state. This uniqueness was proved by
Wald \cite{Wald,Wal78a,Wal78b} who
investigated the criteria that a physically meaningful expectation
value of the stress tensor ought to satisfy.

The back-reaction is formulated in terms of the semiclassical
Einstein equations. These are Einstein equations which have the
expectation value in some quantum state of the stress tensor as a
matter source. The back-reaction problem was investigated in
cosmology, in particular to see whether cosmological anisotropies
could be damped by back-reaction \cite{Lukash,HuPar78}. This was an
earlier attempt \cite{Misner,Mis72} previous to inflation to explain why
the universe is so isotropic at present.

A useful approach to the back-reaction problem is to use the
effective action methods
\cite{Hartle,FisHarHu79,HarHu79} that are familiar in
quantum field theory. These methods were of great help in the
study of cosmological anisotropies since they allowed the
introduction of familiar perturbative treatments into the subject.
The most common effective action method, however, led to equations
of motion which were not real because they were tailored to
compute transition elements of quantum operators rather than
expectation values. Fortunately the appropriate technique had
already been developed by Schwinger and Keldysh
\cite{CTP1,Kel64,ChoEtAl85} in
the so called Closed Time Path (CTP) or ``in-in" effective action
method, and was soon adapted to the gravitational context
\cite{CTP2,CalHu87,Jor87,Paz90}. These techniques were then applied to different
problems in the cosmological context including the effects of
arbitrary perturbations on homogeneous backgrounds \cite{CTP3}. As
a result one was able to deduce the semiclassical Einstein
equations by the CTP functional method: treating the matter fields
as fully quantum fields and the gravitational field
as a classical field.

The semiclassical Einstein equations have limitations even outside
the Planck scales: when the fluctuations around the expectation value
of the stress tensor of the
matter fields are large the semiclassical equations should
break down \cite{Waldfluc,Fordfluc,Kuo and Ford}. One expects, in
fact, that a better approximation would describe the gravitational
field in a probabilistic way. In other words, that the
semiclassical equations should be substituted by some
Langevin-type equations with a stochastic source that describes
the quantum fluctuations. A significant step in this direction was
made by Hu \cite{Hu 89} who proposed to view the back-reaction
problem in the light of the open quantum system paradigm, where
the quantum fields play the role of the ``environment" and the
gravitational field plays the role of the ``system". Following
this proposal a systematic study of the connection between
semiclassical gravity and open quantum systems resulted in the
development of a framework where different semiclassical
Einstein-Langevin equations were derived
\cite{SEL1,HuSin95,HuMat95,CamVer96}. The key
technical factor to most of these results was the use of the
influence functional method of Feynman and Vernon \cite{Feynman,Fey65}
for the description of the system-environment interaction when
only the state of the system is of interest. The CTP method for
open systems involves, in fact, the influence functional.

However although several Einstein-Langevin equations were derived,
these were always partial and related to some particular
cosmological situation. On the other hand, the results were
somewhat formal and some concern could be raised on the physical
reality of the solutions of the stochastic equations for the
gravitational field. This is related to the issue of the
environment induced quantum to classical transition. In fact, for
the existence of a semiclassical regime for the dynamics of the
system one needs two requirements, in the language of the
consistent histories formulation of quantum mechanics
\cite{Griffiths,Omn92}. The first is decoherence, which guarantees that
probabilities can be consistently assigned to histories describing
the evolution of the system, and the second is that these
probabilities should peak near histories which correspond to
solutions of classical equations of motion. The effect of the
environment is crucial on the one hand to provide decoherence
\cite{Zurek} and on the other hand  to produce both dissipation
and noise to the system through back-reaction, thus inducing a
semiclassical stochastic dynamics on the system. As shown by
Gell-Mann and Hartle \cite{Gell-Mann} in an open quantum system
stochastic semiclassical equations are obtained after a coarse
graining of the environmental degrees of freedom and a further
coarse graining in the system variables. That this mechanism could
also work for decoherence and classicalization of the metric field
was not so clear lacking a full quantum description of the
gravitational field, and the analogy could be made only formally
\cite{Martin 99a}.

An alternative axiomatic approach to the Einstein-Langevin
equations which was independent of the open system analogy was
introduced: it was based on the formulation of a general and
consistent dynamical set of equations for a perturbative
correction to semiclassical gravity able to account for the lowest
order quantum stress tensor fluctuations of matter fields
\cite{Martin 99b}. It was later shown that these same equations
could be derived, in this general case, from the influence
functional of Feynman and Vernon in which, the gravitational field
is treated as a classical field and the quantum fields are quantized, the
first being, in fact, the ``system" and the seconds the
``environment" \cite{Martin 99c}. Also, inspired by results in
some simple open quantum systems \cite{CalRouVer03} and results of
stochastic gravity in Minkowski spacetime \cite{Martin 00}, the
concern on the reality of the stochastic solutions was latter
clarified. It was realized that the correlation functions of the
metric fluctuations obtained in stochastic gravity reproduce the
correlation functions in the quantum theory of gravity interacting
with $N$ quantum fields to leading order in an $1/N$ expansion
\cite{HuRouVer04,RouVer06}. Thus, stochastic gravity may be
understood as a powerful and useful framework to study quantum
metric fluctuations.

Here we review the development of stochastic gravity and some of
its applications. In section \ref{2} a brief sketch of
semiclassical gravity is given. In section \ref{3} the axiomatic
approach to stochastic gravity is discussed by introducing the
Einstein-Langevin equations. In section \ref{largeN} to illustrate
the relation between the semiclassical, stochastic and quantum
theories, a simplified model of scalar gravity interacting with
$N$ scalar fields is considered. In section \ref{4} we review an
important application of stochastic gravity in cosmology. It
concerns the computation of the two-point correlations of the
metric perturbations induced by the fluctuation in the stress
tensor of the inflaton field during inflation. The results to
linear order agree with the standard results but the present
method, in which the matter fields are treated exactly, may go
beyond the usual approaches where the inflaton fluctuations are
treated at linear level only. In section \ref{black holes} we deal
with another important application in black hole physics: the
study of the fluctuations near the horizon of an evaporating black
hole. This subject is still under consideration and we only sketch
some of the recent results. Finally, in section \ref{5} we
summarize our result and briefly discuss other applications. We
should mention that two reviews of stochastic gravity and its
applications are now available \cite{HuVer03,HuVer04}.

\section{Semiclassical gravity}
\label{2}

Semiclassical gravity is a theory which describes the interaction
of the gravitational field assumed to be a classical field with
matter fields which are quantum. It is supposed to be some limit
of the still unknown quantum theory describing the interaction of
gravity with other fields. Due to the lack of the full quantum theory,
the semiclassical limit cannot be rigorously derived. However, it
can be formally derived in several ways. One of them is the
leading-order $1/N$ approximation of quantum gravity \cite{Hartle
and Horowitz}, where $N$ is the number of independent free quantum
fields which interact with gravity only. In this limit, after path
integration one arrives at a theory in which the gravitational
field can be treated as a c-number (i.e. is quantized at tree
level) and the quantum fields are fully quantized. If we call
$g_{ab}$ the metric tensor and $\hat\phi$ the scalar field (for
simplicity we consider just one scalar field) one arrives at the
semiclassical Einstein equation as the dynamical equation for the
metric $g_{ab}$:
\begin{equation}
G_{ab}[g]=\kappa \langle \hat T_{ab}[g]\rangle_{ren}, \label{1.1}
\end{equation}
where $\kappa=8\pi G=8\pi/m_P^2$, $G$ is Newton's gravitational
constant and $m_P$ is the Planck mass, $\hat
T_{ab}=T_{ab}[\hat\phi^2]$ is the stress tensor operator which is
quadratic in the field operator $\hat \phi$. This operator, being
the product of distribution valued operators, is ill defined and
needs to be regularized and renormalized, the subscript $ren$
means that the operator has been renormalized. The angle brackets
on the right hand side mean that the expectation value of the
stress tensor operator is computed in some quantum state, say
$|\psi\rangle$, compatible with the geometry described by the
metric $g_{ab}$. On the left hand side $G_{ab}[g]$ stands for the
Einstein tensor for the metric $g_{ab}$ together with the
cosmological constant term and other terms quadratic in the
curvature which are generally needed to renormalize the stress
tensor operator. The quantum field operator $\hat\phi$ propagates
in the background defined by the metric $g$, it thus satisfies a
Klein-Gordon equation,
\begin{equation}
\left(\nabla^2_g-m^2-\xi R[g]\right) \hat\phi=0, \label{1.2}
\end{equation}
where $\nabla^2_g$ stands for the D'Alambert operator in the
background $g_{ab}$, $\xi$ is a dimensionless coupling parameter
($\xi=0$ is the minimal coupling and $\xi=1/6$ is the conformal
coupling) and $R$ is the Ricci scalar for the background metric.
Equation (\ref{1.1}) is the semiclassical Einstein equation, it is
the dynamical equation for the metric tensor $g_{ab}$ and
describes the back-reaction of the quantum matter fields on the
geometry. A solution of semiclassical gravity consists of the set
$(g_{ab},\hat \phi,|\psi\rangle)$ where $g_{ab}$ is a solution of
(\ref{1.1}), $\hat \phi$ is a solution of (\ref{1.2}) and
$|\psi\rangle$ is the quantum state in which the expectation value
of the stress tensor in equation (\ref{1.1}) is computed.

This theory is in some sense unique as a theory where the
gravitational field is classical. In fact, the (classical)
gravitational field interacts with matter fields through the
stress tensor, and the only reasonable c-number stress tensor that
one may construct \cite{Wald} with the operator $\hat T_{ab}$ is
just the right hand side of (\ref{1.1}), modulo the curvature
terms needed for renormalization. However the scope and limits of
the theory are not so well understood as a consequence of the lack
of the full quantum theory. It is assumed that the semiclassical
theory should break down at Planck scales, which is when simple
order of magnitude estimates suggest that the quantum effects of
gravity cannot be ignored: the gravitational energy of a quantum
fluctuation of energy in a Planck size region, determined by the
Heisenberg uncertainty principle, is of the same order of
magnitude as the energy fluctuation.

There is also another situation when the semiclassical theory
should break down, namely, when the fluctuations of the stress
tensor are large. This has been emphasized by Ford and
collaborators. It is less clear how to quantify what a large
fluctuation here means and some criteria have been proposed
\cite{Kuo and Ford,Phillips and Hu,PhiHu00}. Generally this depends on the
quantum state and may be illustrated by the example used in ref.
\cite{Fordfluc} as follows.

Let us assume a quantum state formed by an isolated system which
consists of a superposition with equal amplitude of one
configuration with mass $M_1$ and another with mass $M_2$.
Semiclassical theory as described in (\ref{1.1}) predicts that the
gravitational field of this system is produced by the average mass
$(M_1+M_2)/2$, that is a test particle will move on the background
spacetime produced by such a source. However one would expect that
if we send a succession of test particles to probe the
gravitational field of the above system, half of the time they
would react to the field of a mass $M_1$ and the other half to the
field of a mass $M_2$. If the two masses differ substantially the
two predictions are clearly different, note that the fluctuations
in mass of the quantum state is of the order of $(M_1-M_2)^2$.
Although the previous example is suggestive a word of caution
should be said in order not to take it too literally. In fact, if
the previous masses are macroscopic the quantum system decoheres
very quickly \cite{Zurek} and instead of a pure quantum state it
is described by a density matrix which diagonalizes in a certain
pointer basis. For observables associated to this pointer basis
the matrix density description is  equivalent to that provided by
a statistical ensemble. In any case, however, from the point of
view of the test particles the predictions differ from that of the
semiclassical theory.

\section{Stochastic gravity}
\label{3}

The purpose of semiclassical stochastic gravity, or stochastic
gravity for short, is to be able to deal with the situation of the
previous example in which the predictions of the semiclassical
theory may be too rough. Consequently, our first point is to
characterize the quantum fluctuations of the stress tensor.

The physical observable that measures these fluctuations is
related to the two-point stress tensor correlations. Let us
consider the tensor operator $\hat t_{ab}\equiv \hat T_{ab}-
\langle \hat T_{ab}\rangle\hat I$, where $\hat I$ is the identity
operator, and introduce the {\it{noise kernel}} as the four index
bi-tensor defined as the expectation value of the anticommutator
of the operator $\hat t_{ab}$:
\begin{equation}
N_{abcd}(x,y)=\frac{1}{2}\langle\{ \hat t_{ab}(x),\hat t_{cd}(y)\}
\rangle. \label{1.3}
\end{equation}
This expectation value is taken in the background metric $g_{ab}$
which we assume to be a solution of the semiclassical equation
(\ref{1.1}). An important property of the symmetric bi-tensor,
$N_{abcd}(x,y)=N_{cdab}(y,x)$, is that it is finite because the
tensor operator $\hat t_{ab}$ is finite since the ultraviolet
divergences of $\hat T_{ab}$ are cancelled by the substraction of
$\langle \hat T_{ab}\rangle$. Since the operator $\hat T_{ab}$ is
selfadjoint $N_{abcd}(x,y)$, which is the expectation value of an
anticommutator, is real and positive semi-definite. This last
property allows for the introduction of a classical Gaussian
stochastic tensor $\xi_{ab}$ defined by
\begin{equation}
\langle\xi_{ab}(x)\rangle_s=0,\ \ \
\langle\xi_{ab}(x)\xi_{cd}(y)\rangle_s=N_{abcd}(x,y). \label{1.4}
\end{equation}

This stochastic tensor is symmetric $\xi_{ab}=\xi_{ba}$ and
divergenceless, $\nabla^a\xi_{ab}=0$, as a consequence of the fact
that the stress tensor operator is divergenceless. The subscript
$s$ means that the expectation value is just a classical
stochastic average. Note that we assume that $\xi_{ab}$ is
Gaussian for simplicity, in order to include the main effect. The
idea now is simple, we want to modify the semiclassical Einstein
equation (\ref{1.1}) by introducing a linear correction to the
metric tensor $g_{ab}$, such as $g_{ab}+h_{ab}$, which accounts
consistently for the fluctuations of the stress tensor. The
simplest equation is obtained by adding to the right hand side of
equation (\ref{1.1}), but written for the perturbed metric, the
stochastic tensor just introduced. Substraction of the
semiclassical equation (\ref{1.1}) leads to the following
equation,
\begin{equation}
G_{ab}^{(1)}[g+h]=\kappa\langle \hat
T_{ab}^{(1)}[g+h]\rangle_{ren}+\kappa\xi_{ab}[g]),\label{1.5}
\end{equation}
where we recall that the background metric $g_{ab}$ is assumed to
be a solution of equation (\ref{1.1}). As indicated by the
superscript $(1)$ this stochastic equation must be thought of as a
linear equation for the metric perturbation $h_{ab}$ which will
behave consequently as a stochastic field tensor. Note that the
tensor $\xi_{ab}[g]$ is not a dynamical source, since it has been
defined in the background metric. Note also that this source is
divergenceless with respect to the metric, and it is thus
consistent to write it on the right hand side of the Einstein
equation. This equation is gauge invariant with respect to
diffeomorphisms defined by any field on the background spacetime
\cite{Martin 99b}. If we take the statistical average, equation
(\ref{1.5}) becomes just the semiclassical equation for the metric
$g_{ab}+\langle h_{ab}\rangle_s$ where the expectation value of
$\hat T_{ab}$ is taken in the perturbed spacetime. The quantum
field now propagates in the spacetime described by the perturbed
metric and thus it satisfies
\begin{equation}
\left(\nabla^2_{g+h}-m^2-\xi R[g+h]\right) \hat\phi=0, \label{1.5a}
\end{equation}
where the Ricci scalar is evaluated for the perturbed metric.

The stochastic equation (\ref{1.5}) is known as the
Einstein-Langevin equation. The equation predicts that the
gravitational field has stochastic fluctuations over the
background $g_{ab}$. It is linear in $h_{ab}$, thus its solutions
can be written as follows,
\begin{equation}
h_{ab}(x)=h_{ab}^0(x)+\kappa\int d^4x^\prime
\sqrt{-g(x^\prime)}G_{abcd}^{ret}(x,x^\prime)\xi^{cd}(x^\prime),
\label{1.5b}
\end{equation}
where $h_{ab}^{0}(x)$ is the solution of the homogeneous equation
containing information on the initial conditions and
$G_{abcd}^{ret}(x,x^\prime)$ is the retarded propagator of
equation (\ref{1.5}) with vanishing initial conditions. Form this
we obtain the two-point correlations for the metric perturbations:
\begin{equation}
\langle h_{ab}(x)h_{cd}(y)\rangle_s = \langle
h_{ab}^0(x)h_{cd}^0(y)\rangle_s +\kappa^2\int d^4x^\prime
d^4y^\prime \sqrt{g(x^\prime)g(y^\prime)}
G_{abef}^{ret}(x,x^\prime) N^{efgh}(x^\prime,y^\prime)
G_{cdgh}^{ret}(y,y^\prime). \label{1.5c}
\end{equation}
There are two different contributions to the two-point
correlations. The first one is connected to the fluctuations of
the initial state of the metric perturbations, we refer to them as
\textit{intrinsic fluctuations}. The second contribution is
proportional to the noise kernel and is thus connected with the
fluctuations of the quantum fields, we will refer to them as
\textit{induced fluctuations}. These two-point stochastic
correlations are the most relevant physical observable, to find
them requires to know the noise kernel $N_{abcd}(x,y)$. Note that
the noise kernel should be thought of as a distribution function,
the limit of coincidence points has meaning only in the sense of
distributions. Explicit expressions of this kernel in terms of the
two point Wightman functions is given in \cite{Martin 99b},
expressions based on point-splitting methods have also been given
in \cite{Roura 00a,Phillips and Hu 00}.

These stochastic correlations for the metric perturbations satisfy
a very important property. In fact, it can be shown that they
correspond exactly to the symmetrized two-point quantum metric
correlations obtained in the large $N$ expansion:
\begin{equation}
\frac{1}{2}\langle\{\hat h_{ab}(x),\hat h_{cd}(y)\}\rangle =
\langle h_{ab}(x)h_{cd}(y)\rangle_s, \label{1.5d}
\end{equation}
where $\hat h_{ab}(x)$ mean the quantum operator corresponding to
the metric perturbations.  This result was implicitly obtained in
the Minkowski background in ref. \cite{Martin 00} where the
two-point correlation in the stochastic context was computed for
the linearized metric perturbations. This stochastic correlation
exactly agrees with the symmetrized part of the graviton
propagator computed by Tomboulis \cite{Tom77} in the quantum
context of gravity interacting with $N$ matter Fermion fields,
where the graviton propagator is of order $1/N$. This result can
be extended to an arbitrary background in the context of the large
$N$ expansion \cite{RouVer06}, a sketch of the proof with explicit
details in the Minkowski background can be found in ref.
\cite{HuRouVer04}. This connection between the stochastic
correlations and the quantum correlations was noted and studied in
detail in the context of simpler open quantum systems
\cite{CalRouVer03}. It is thus clear that stochastic gravity goes
beyond semiclassical gravity in the following sense. The
semiclassical theory, which is based on the expectation value of
the stress energy tensor, carries information on the field
two-point correlations only, since $\langle \hat T_{ab}\rangle$ is
quadratic in the field operator $\hat\phi$. The stochastic theory
on the other hand, is based on the noise kernel (\ref{1.3})  which
is quartic in the field operator. However, it does not carry
information on the graviton-graviton interaction, which in the
context of the large $N$ expansion it gives diagrams of order
$1/N^2$. This will be illustrated in section \ref{largeN}.

\subsection{Functional approach}

To end this section we should mention that the Einstein-Langevin
equation (\ref{1.5}) may also be derived using the CTP functional
method \cite{Martin 99a}. As remarked in the introduction the CTP
functional was introduced by Schwinger
\cite{CTP1,Kel64,ChoEtAl85} to compute
expectation values. One just considers the interaction of the
gravitational field $g_{ab}$, classical, with the field
$\phi$, fully quantum. Then the effective action for the
gravitational field is derived after integrating out the degrees
of freedom of the quantum field, and the CTP influence action
reduces basically to the Feynman and Vernon influence functional
\cite{Feynman,Fey65} used in quantum open systems. Here the system is
the gravitational field and the environment is the quantum field.
The stochastic terms for the gravitational field are found by
suitably interpreting some pure imaginary term which appear in the
influence action. These terms are closely connected to Gell-Mann
and Hartle decoherence functional \cite{Gell-Mann} used to study
decoherence and classicalization in open quantum systems. The net
result is that the interaction with the environment induces
fluctuations in the system dynamics. It is precisely the noise
kernel introduced in (\ref{1.3}) that accounts for this effect.

\section{The large $N$ expansion}
\label{largeN}

The large $N$ expansion has been successfully used in quantum
cromodynamics to compute some non-perturbative results. This
expansion re-sums and rearranges Feynman perturbative series
including self-energies. For gravity interacting with $N$ matter
fields it shows that graviton loops are of higher order than matter
loops. To illustrate the large $N$ expansion let us consider the
following toy model of gravity, which we will simplify as a scalar
field $h$, interacting with a scalar field $\phi$ described by the
Lagragian density
\begin{equation}
L=\frac{1}{\kappa}\int d^4 x\left(\partial_a h\partial^a
h+h(\partial h)^2+\dots\right)-\int d^4 x\left(\partial_a \phi\partial^a
\phi+m^2\phi^2\right)+\int d^4 x\left( h(\partial\phi)^2+\dots\right),
 \label{N1}
\end{equation}
where, as previously $\kappa=8\pi G$, and we have assumed that the
interaction is linear in the (dimensionless) scalar gravitational
field $h$ and quadratic in the matter field $\phi$ to simulate in
a simplified way the coupling of the metric with the stress tensor
ofº the matter fields.
We have also included a self coupling graviton term of $O(h^3)$
which also appears in perturbative gravity beyond the linear
approximation.

We may now compute the dressed graviton propagator perturbatively
as the following series of Feynman diagrams. The first diagram is
just the free graviton propagator which is of $O(\kappa)$, as one
can see from the kinetic term for the graviton in equation
(\ref{N1}). The next diagram is one loop of matter with two
external legs which are the graviton propagators. This diagram has
two vertices with one graviton propagator and two matter field
propagators. Since the vertices and the matter propagators
contribute with 1 and each graviton propagator contributes with a
$\kappa$ this diagram is of order $O(\kappa^2)$. The next diagram
contains two loops of matter and three gravitons, and consequently
it is of order $O(\kappa^3)$. There will also be terms with one
graviton loop and two graviton propagators as external legs, with
three graviton propagators at the two vertices due to the $O(h^3)$
term in the action (\ref{N1}). Since there are four graviton
propagators which carry a $\kappa^4$ but two vertices which have
$\kappa^{-2}$ this diagram is of order $O(\kappa^2)$, like the
term with one matter loop.

Let us now consider the large $N$ expansion. We assume that the
gravitational field is coupled with a large number of identical
fields $\phi_j$, $j=1,\dots,N$ which couple only with $h$. Next we
re-scale the gravitational coupling in such a way that
$\bar\kappa=\kappa N$ is finite even when $N$ goes to infinity.
The Lagrangian density of this system is:
\begin{equation}
L=\frac{N}{\bar\kappa}\int d^4 x\left(\partial_a h\partial^a
h+h(\partial h)^2+\dots\right)-\sum_j^N\int d^4 x\left(\partial_a \phi_j\partial^a
\phi_j+m^2\phi^2\right)+\sum_j^N\int d^4 x\left( h(\partial\phi_j)^2+\dots\right).
 \label{N2}
\end{equation}

Now and expansion in powers of $1/N$ of the dressed graviton
propagator is given by the following series of Feynman diagrams.
The first diagram is the free graviton propagator which is now of
order $O(\bar\kappa/N)$ the following diagrams are $N$ identical
Feynman diagrams with one loop of matter and two graviton
propagators as external legs, each diagram due to the two graviton
propagators is of order $O(\bar\kappa^2/N^2)$ but since there are
$N$ of them the sum can be represented by a single diagram with a
loop of matter of weight $N$, and therefore this diagram is of
order $O(\bar\kappa^2/N)$. This means that it is of the same order
as the first diagram in an expansion in $1/N$. Then there are
diagrams with two loops of matter and three graviton propagators,
as before we can assign a weight of $N$ to each loop and taking
into account the three graviton propagators this diagram is of
order $O(\bar\kappa^3/N)$, and so on. This means that to order
$1/N$ the dressed graviton propagator contains all the
perturbative sums in powers of $\bar\kappa$ of the matter loops.

Next, there is a diagram with one graviton loop and two graviton
legs. Let us count the order of this diagram: it contains four
graviton propagators and two vertices, the propagators contribute
as $(\bar\kappa/N)^4$ and the vertices as $(N/\bar\kappa)^2$, thus
this diagram is of $O(\bar\kappa^2/N^2)$. Therefore graviton loop
contributes to higher order in the $1/N$ expansion than matter
loops. Similarly there are $N$ diagrams with one loop of matter
with an internal graviton propagator and two external graviton
legs. Thus we have three graviton propagators and since there are
$N$ of them, their sum is of order $O(\bar\kappa^3/N^2)$. To
summarize, we have that when $N\to\infty$ there are no graviton
propagators and gravity is classical, this is semiclassical
gravity. To next to leading order, $1/N$, the graviton propagator
includes all matter loop contributions, but no contributions from
graviton loops and internal graviton propagators in matter loops.
This is what stochastic gravity reproduces.

That stochastic gravity is connected to the large $N$ expansion
can be seen from the stochastic correlations of linear metric
perturbations on the Minkowski background computed in ref.
\cite{Martin 00}. These correlations are in exact agreement with
the imaginary part of the graviton propagator found by Tomboulis
in the large $N$ expansion for the quantum theory of gravity
interacting with $N$ Fermion fields \cite{Tom77}. This has been
proved in detail in ref. \cite{HuRouVer04} and extended to the
general case \cite{RouVer06}.

\section{Gravitational fluctuations during inflation}
\label{4}

An important application of stochastic gravity is the derivation
of the cosmological perturbations generated during inflation
\cite{Roura 00a}. Let us consider the Lagrangian density for an
inflaton field $\phi$ of mass $m$
\begin{equation}
{\cal L}(\phi)=\frac{1}{2}\partial_a\phi\partial^a\phi +
\frac{1}{2}m^2\phi^2, \label{1.14}
\end{equation}
which is the basis of the simplest chaotic inflationary model
\cite{Linde 90}. The conditions for the existence of an
inflationary period, which is characterized by an accelerated
expansion of the spacetime, is that the value of the field
averaged over a region with the typical size of the Hubble radius
is higher than the Planck mass $m_P$. After the Planck era one
expects, by Heisenberg uncertainty principle, that in a region of
the size of the Planck size $m_P^{-1}$ the energy density would be
of the order of $m_P^{4}$. If the inflaton field starts in one of
those regions at a value much larger than $m_P$ then with a
potential like that of eq. (\ref{1.14}) one expects an inflaton
mass $m\ll m_P$ in which case the potential is very flat and the
field rolls slowly towards the minimum producing a period of
almost de Sitter expansion of that region. In order to solve the
cosmological horizon and flatness problem more than 60 e-folds of
expansion are needed, to achieve this the scalar field should
begin with a value higher than $3m_P$. Moreover, we will see that
the large scale anisotropies measured
\cite{Smoot 92,Ben03,Pei03} restrict the
inflaton mass to be of the order of $10^{-6}m_P$.

We want to study the metric perturbations produced by the stress tensor
fluctuations of the inflaton field on the homogeneous background of a flat
Friedmann-Robertson-Walker model, described by the cosmological scale
factor $a(\eta)$, where $\eta$ is the conformal time, which is driven by
the homogeneous inflaton field $\phi(\eta)=\langle\hat\phi\rangle$. Thus we
write the inflaton field in the following form
\begin{equation}
\hat\phi=\phi(\eta)+ \hat\varphi (x),
\label{1.15}
\end{equation}
where $\hat\varphi (x)$ corresponds to a free massive quantum
scalar field with zero expectation value on the homogeneous
background metric: $\langle\hat\varphi\rangle=0$. Restricting
ourselves to scalar-type perturbations the perturbed metric
$\tilde g_{ab}=g_{ab}+h_{ab}$ can be written in the longitudinal
gauge as,
\begin{equation}
ds^2=a^2(\eta)[-(1+2\Phi(x))d\eta^2+(1-2\Psi(x))\delta_{ij}dx^idx^j],
\label{1.16}
\end{equation}
where the metric perturbations $\Phi(x)$ and $\Psi(x)$ correspond to
Bardeen's gauge invariant variables \cite{Bardeen,Mukhanov 92}.

The Einstein-Langevin equation is gauge invariant, and thus we can
work in a desired gauge and then extract the gauge invariant
quantities. The Einstein-Langevin equation (\ref{1.5}) reads now:
\begin{equation}
G_{ab}^{(1)}[h]-\kappa\langle\hat T_{ab}^{(1)}[h]\rangle=
\kappa\xi_{ab}[g], \label{1.17}
\end{equation}
where $g_{ab}$ satisfies the semiclassical Einstein equations;
assuming slow roll this background metric is an almost de Sitter
metric. The superscript $(1)$ refers to functions linear in the
metric perturbation $h_{ab}$. The stress tensor operator $\hat
T_{ab}$ for the minimally coupled inflaton field in the perturbed
metric $\tilde g_{ab}$ is:
\begin{equation}
\hat T_{ab}=\tilde\nabla_{a}\hat\phi\tilde\nabla_{b}\hat\phi+
\frac{1}{2}\tilde g_{ab}
(\tilde\nabla_c\hat\phi\tilde\nabla^c\hat\phi+ m^2\hat\phi^2).
\label{1.18}
\end{equation}

Now, using the decomposition of the scalar field into its
homogeneous and inhomogeneous part, see eq. (\ref{1.15}), and the
metric $\tilde g_{ab}$ into its homogeneous background $g_{ab}$ and
its perturbation $h_{ab}$, the renormalized expectation value for
the stress tensor can be written as
\begin{equation}
\langle \hat T_{ab}[\tilde g]\rangle= \langle \hat T[\tilde
g]\rangle_{\phi\phi}+ \langle \hat T_{ab}[\tilde
g]\rangle_{\phi\varphi}+ \langle \hat T_{ab}[\tilde
g]\rangle_{\varphi\varphi}, \label{1.19}
\end{equation}
where only the homogeneous solution for the scalar field
contributes to the first term. The second term is proportional to
$\langle\hat\varphi[\tilde g]\rangle$ which is not zero because
the field dynamics is considered on the perturbed spacetime, i.e.
this term includes the coupling of the field with $h_{ab}$, as
seen in equation (\ref{1.5a}). The last term corresponds to the
expectation value to the stress tensor for a free scalar field on
the spacetime of the perturbed metric.

We can now compute the noise kernel $N_{abcd}(x,y)$ defined in
equation (\ref{1.3}), which after using the previous decomposition
may be written as
\begin{equation}
\langle \{\hat t_{ab},\hat t_{cd}\}\rangle[g]= \langle \{\hat
t_{ab},\hat t_{cd}\}\rangle_{\phi\varphi}[g]+ \langle \{\hat
t_{ab},\hat t_{cd}\}\rangle_{\varphi\varphi}[g], \label{1.20}
\end{equation}
where we have used the fact that $\langle\hat\varphi\rangle=0
=\langle\hat\varphi\hat\varphi\hat\varphi\rangle$ for Gaussian
states on the background geometry. We have considered the vacuum
state to be the Bunch-Davies, or Euclidean, vacuum which is
preferred in the de Sitter background, and this state is Gaussian.
In the above equation the first term is quadratic in $\hat\varphi$
whereas  the second one is quartic, both contributions to the
noise kernel are separately conserved since both $\phi(\eta)$ and
$\hat\varphi$ satisfy the Klein-Gordon field equations on the
background spacetime. Consequently, the two terms can be
considered separately. On the other hand if one treats $\hat
\varphi$ as a small perturbation the second term in (\ref{1.20})
is of lower order than the first and may be neglected
consistently, this corresponds to neglecting
the fluctuations associated with the last term in equation
(\ref{1.19}). This approximation is equivalent to keep only linear
terms in the inflaton perturbations. Stress tensor fluctuations
due to a term like the last term of (\ref{1.19}) were considered
in ref. \cite{Roura 99}.

We can now write down the Einstein-Langevin equations (\ref{1.17}). It is
easy to check that the {\it space-space} components coming from the stress
tensor expectation value terms and the stochastic tensor are diagonal,
i.e. $\langle\hat T_{ij}\rangle=0= \xi_{ij}$ for $i\not= j$. This, in
turn, implies that the two functions characterizing the scalar metric
perturbations are equal: $\Phi=\Psi$ in agreement with ref.
\cite{Mukhanov 92}. The equation for $\Phi$ can be obtained from the
$0i$-component of the Einstein-Langevin equation, which in Fourier
space reads
\begin{equation}
2ik_i(H\Phi_k+\Phi'_k)= \frac{8\pi}{m_P^2}\xi_{k\, 0i},
\label{1.21}
\end{equation}
where $k_i$ is the comoving momentum component associated to the
comoving coordinate $x^i$. Here primes denote derivatives with
respect to the conformal time $\eta$ and $H=a'/a$. A non-local
term of dissipative character which comes from the second term in
(\ref{1.19}) should also appear on the left hand side of equation
(\ref{1.21}), but we have ignored this term for simplicity (if one
includes this non-local term it is then  more convenient to write
an equation combining the other equations which is free of
non-local terms; but the results are not substantially altered).
To solve this equation, whose left hand side comes from the
linearized Einstein tensor for the perturbed metric \cite{Mukhanov
92}, we need the retarded propagator for the gravitational
potential $\Phi_k$,
\begin{equation}
G_k(\eta,\eta')= -i \frac{4\pi}{k_i m_P^2}\left( \theta(\eta-\eta')
\frac{a(\eta')}{a(\eta)}+f(\eta,\eta')\right),
\label{1.22}
\end{equation}
where $f$ is a homogeneous solution of (\ref{1.21}) related to the
initial conditions chosen. For instance, if we take
$f(\eta,\eta')=-\theta(\eta_0-\eta')a(\eta')/a(\eta)$ the solution
would correspond to ``turning on" the stochastic source at
$\eta_0$.

The correlation function for the metric perturbations is now given
by
\begin{equation}
\langle\Phi_k(\eta)\Phi_{k'}(\eta')\rangle_s= (2\pi)^2\delta(\vec
k+\vec k')\int^\eta d\eta_1\int^{\eta'}d\eta_2 G_k(\eta,\eta_1)
G_{k'}(\eta',\eta_2) \langle\xi_{k\ 0i}(\eta_1)\xi_{k'\
0i}(\eta_2)\rangle_s . \label{1.23}
\end{equation}
The correlation function for the stochastic source , which is connected to
the stress tensor fluctuations through the noise kernel is given by,
\begin{equation}
\langle\xi_{k\ 0i}(\eta_1)\xi_{-k\ 0i}(\eta_2)\rangle_s= \frac{1}{2}
\langle\{\hat t^k_{0i}(\eta_1,\hat
t^{-k}_{0i}(\eta_2)\}\rangle_{\phi\varphi}= \frac{1}{2}
k_ik_i\phi'(\eta_1)\phi'(\eta_2)G_k^{(1)}(\eta_1,\eta_2),
\label{1.24}
\end{equation}
where $G_k^{(1)}(\eta_1,\eta_2)=\langle\{\hat\varphi_k(\eta_1),
\hat\varphi_{-k}(\eta_2)\}\rangle$ is the $k$-mode Hadamard function for a
free minimally coupled scalar field which is in the
Euclidean vacuum on the de Sitter background.

It is useful to compute the Hadamard function for a massless field and
consider a perturbative expansion in terms of the dimensionless parameter
$m/m_P$. Thus we consider $\bar
G_k^{(1)}(\eta_1,\eta_2)= a(\eta_1)a(\eta_2)G_k^{(1)}(\eta_1,\eta_2)=
\langle 0|\{\hat y_k(\eta_1),\hat y_{-k}(\eta_2)\}|0\rangle=
2{\cal R}\left(u_k(\eta_1)u_k^*(\eta_2)\right)$
with $\hat y_k(\eta)= a(\eta)\hat\varphi_k(\eta)=
\hat a_k u_k(\eta)+\hat a_{-k}^\dagger u_{-k}^*(\eta)$ and where
$u_k=(2k)^{-1/2}e^{ik\eta}(1-i/\eta)$ are the positive
frequency $k$-mode for a massless minimally coupled scalar field
on a de Sitter background, which define the
Euclidean vacuum state: $\hat a_k|0\rangle=0$ \cite{Birrell 84}.

The background geometry, however, is not exactly that of de Sitter
spacetime, for which $a(\eta)=-(H\eta)^{-1}$ with $-\infty <\eta< 0$.
One can expand in terms of the ``slow-roll" parameters and assume that to
first order $\dot\phi(t)\simeq m_P^2(m/m_P)$, where $t$ is the physical
time. The correlation function for the metric perturbation
(\ref{1.23}) is the computed, see ref. \cite{Roura 00a} for details. The
final result, however, is very weakly dependent on the initial conditions
as one may understand from the fact that the accelerated expansion of de
quasi-de Sitter spacetime during inflation erases the information about
the initial conditions. Thus one may take the initial time to be
$\eta_0=-\infty$ and obtain to lowest order in $m/m_P$ the expression
\begin{equation}
\langle\Phi_k(\eta)\Phi_{k'}(\eta')\rangle_s\simeq 8\pi^2\left(
\frac{m}{m_P}\right)^2 k^{-3}(2\pi)^3\delta(\vec k+\vec k') \cos
k(\eta-\eta'). \label{1.25}
\end{equation}

{}From this result two main conclusions are derived. First, the
prediction of an almost Harrison-Zel'dovich scale-invariant
spectrum for large scales, i.e. small values of $k$. Second, since
the correlation function is of order of $(m/m_P)^2$ a severe bound
to the mass $m$ is imposed by the gravitational fluctuations
derived from the small values of the Cosmic Microwave Background
(CMB) anisotropies detected by COBE \cite{Smoot 92}
and WMAP \cite{Ben03,Pei03}.
This bound is of the order of
$(m/m_P)\sim 10^{-6}$ \cite{Smoot 92,Mukhanov 92}. One possible
advantage of the Einstein-Langevin approach to the gravitational
fluctuations in inflaton over the approach based on the
quantization of the linear perturbations of both the metric and
the inflaton field \cite{Mukhanov 92}, is that an exact treatment
of the inflaton quantum fluctuations is possible. On the other
hand although the gravitational fluctuations are here assumed to
be classical, the correlation functions obtained correspond to the
expectation values of the quantum metric perturbations, as we have
remarked in the previous section \cite{CalRouVer03,RouVer06}.

\section{Fluctuations near black hole horizons}
\label{black holes}

Another interesting application of stochastic gravity is found in
the context of black hole physics, in particular the stress tensor
fluctuations near the black hole horizon may induce fluctuations
in the horizon area. The relevance of these fluctuations in
Hawking radiation needs to be understood \cite{BarFroPar00}. Some
preliminary investigations seem to indicate that the fluctuations
of the black hole horizon are always small and that the Hawking
result should not be substantially different \cite{Ford2,Ford3},
however, some other results by Bekenstein \cite{Bekenstein} seem
to point in the opposite direction suggesting that the
fluctuations of the black hole horizon may be significant in the
long run. The contribution of the horizon fluctuations to the
black hole entropy \cite{Sorkin,SorSud99} is another interesting issue that
may deserve some attention in the present context.

To clarify this situation Hu and Roura \cite{HuRou06} have
analyzed this back-reaction problem in the stochastic gravity
framework. Due to technical difficulties this is still work in
progress so I will only summarize very briefly the main results.
As shown by Hawking a black hole formed by spherical collapse
emits thermal radiation with a temperature $T=m_P^2/8\pi M$, where
$M$ is the mass of the black hole. This calculation was made under
the assumption of quantum field theory in a curved background.
That is, assuming that the black hole has a fixed mass much larger
than the Planck mass $m_P$, so that one can safely ignore quantum
gravity effects, and that the black hole exterior can be described
by the Schwarzschild metric. But, energy conservation arguments
indicate that as the black hole emits radiation it will loss mass
and evaporate. A precise calculation of the evaporation process
requires the use of the semiclassical Einstein equations which
describes the back-reaction of the quantum matter fields on the
gravitational field in a self-consistent way. An exact
self-consistent calculation of the evaporation process is by no
means easy and has not been performed. Note that the black hole
exterior is not vacuum any more, as the expectation value of the
stress tensor $\langle \hat T_{ab}\rangle_{ren}$ is not zero, and
is not described exactly by the Schwarzschild metric. Thus, even
the radiation process needs to be reviewed. Furthermore even in
the Schwarzschild background an exact analytic expression to describe
that expectation value in the Unruh vacuum, which is the natural
quantum state that describes the initial vacuum in a black hole
formed by gravitational collapse, is not known.

Fortunately, for large black holes the evaporation process is slow
and a quasi-adiabatic approximation can be used to solve Einstein
semiclassical equations. In this approximation one can assume that
the black hole exterior is described by a Schwarzschild metric
with a mass $M$ which is the mass that the black hole has at that
time. A suitable parameter to use in this approximation is the
luminosity at a given time $L_H=B/M^2$, where $B$ is a constant
which depends on the number of fields considered, the spins of
these fields and the grey body factor. It has been estimated by
Page \cite{Pag76} to be of the order of $10^{-4}$. Here, and in
the rest of this section, we use units in which $m_P=1$. The
quasi-adiabatic approximation holds as long as $L_H\ll 1$. Black
hole evaporation in the adiabatic approximation was described by
Bardeen \cite{Bar81} and Massar \cite{Mas95} for spherically
symmetric black holes. Let us summarize this calculation. A
spherically symmetric metric can always be written as
\begin{equation}
ds^2=-e^{\psi(v,r)}\left[1-\frac{2m(v,r)}{r}\right]dv^2+
2e^{\psi(v,r)}dv\,dr +r^2(d\theta^2+\sin^2\theta d\varphi^2).
 \label{bh1}
\end{equation}
There is an apparent horizon where the expansion of the outgoing
null geodesics vanish, that is where $dr_a/dv=0$, which leads to
$r_a(v)=2m(v,r_a(v))\equiv 2M(v)$. Thus we have defined $2M(v)$ as
the apparent horizon, when $M$ is constant this corresponds to the
event horizon and $M$ is the black hole mass. Thus in the
adiabatic approximation we may consider that $M(v)$ is the black
hole mass at the ``advanced" time $v$. The semiclassical Einstein
equations become in the above coordinates,
\begin{equation}
\frac{\partial m}{\partial v}=4\pi r^2\langle T_v^r\rangle,\ \ \
\frac{\partial m}{\partial r}=-4\pi r^2\langle T_v^v\rangle,\ \ \
\frac{\partial \psi}{\partial v}=4\pi r^2\langle T_{rr}\rangle.
 \label{bh2}
\end{equation}
We do not have an analytic expression for the expectation value of
the stress tensor even in the Schwarzschild spacetime. However, at
large radii it corresponds to a thermal flux of radiation and we
may write $\langle T_v^r\rangle=L_H/(4\pi r^2)$. Then one can use
the stress tensor conservation equation to relate components on
the horizon and far from it,
\begin{equation}
\frac{\partial(r^2\langle T_v^r\rangle) }{\partial r}+
r^2\frac{\partial\langle T_v^v\rangle }{\partial v}=0.
 \label{bh3}
\end{equation}
Using this equation one may relate the positive energy flux
radiated away from the horizon and the negative energy flux
crossing the horizon. Taking this relation into account and the
quasi-adiabatic approximation one finally gets, from the first of
equations ({\ref{bh2}}) at the horizon, the equation for the
evolution of the apparent horizon:
\begin{equation}
\frac{dM }{dv}=-\frac{B }{M^2},
 \label{bh3a}
\end{equation}
which gives the evaporation rate, as one would expect from energy
conservation considerations.

Now the Einstein-Langevin equation may be used to study the metric
fluctuations near the black hole horizon of the evaporating black
hole. A full self-consistent computation is technically involved,
in particular the computation of the noise kernel is very
complicated, but Roura and Hu \cite{HuRou06} where able to give
some reasonable estimates of the event horizon fluctuations in the
evaporating process. They concentrate on the spherically symmetric
fluctuations by projecting the Einstein-Langevin equation on the
spherical sector. The Einstein-Langevin equation for the
perturbation of $m(v.r)$, $\delta m(v,r)$, can be written in the
quasi-adiabatic approximation as
\begin{equation}
\frac{\partial(\delta m)}{\partial v}=\frac{2B}{m^3}\delta m +4\pi
r^2\xi_v^r +O(L_H^2), \label{bh4}
\end{equation}
where $\xi_v^r$ is the Gaussian stochastic component defined
through the noise kernel components $\frac{1}{2}\langle \{\hat
t_v^r\hat t_v^r\}\rangle$. Here one is interested in the
fluctuations near the apparent horizon. The noise kernel there has
not yet been computed, however, far from the horizon it has been
estimated by Wu and Ford \cite{Ford3}. They found a correlation
time for the fluctuations of the stress tensor of the order of $M$
and smearing the 2-point functions over this correlation time they
found fluctuations of the order of $1/M^4$. Form this Hu and Roura
\cite{HuRou06} deduced that
\begin{equation}
\langle\xi(v)\xi(v^\prime)\rangle_s\sim\frac{1}{M^3(v)}\delta(v-v^\prime),
\label{bh5}
\end{equation}
which for times larger than the correlation time reproduces the
Wu-Ford result \cite{Ford3}. Here $\xi(v)\equiv (4\pi
r^2\xi_v^r)[v,r\sim 6M(v)]$, relatively far from the horizon.

One then assumes that as a consequence of the divergenceless
property of the stochastic tensor which leads to an equation like
(\ref{bh2}), but where $\xi_a^b$ replaces $\langle T_a^b\rangle$,
one may be able to connect in a simple way the value of the
stochastic source near the horizon with its value at large radii.
Then, under the assumption that equation (\ref{bh5}) is still
valid at the horizon Hu and Roura \cite{HuRou06} found that if
$M_0$ is the initial mass of the black hole the fluctuations of
$M(v)$ are,
\begin{equation}
\delta M\sim\left(\frac{M_0}{M}\right)^2,
\label{bh6}
\end{equation}
which imply that the fluctuations grow in time and they become of
the order of the mean value $\delta M(v)\sim M(v)$ when $M(v)\sim
M_0^{2/3}$. From the evaporation rate equation (\ref{bh3a}) the
evaporation time of the black hole is of order of $M_0^3$ and thus
it reaches the mass $M_0^{2/3}$ after a significant fraction of
the evaporation time. For instance for a black hole with an
initial mass of the order the solar mass $M_0=1\,M_{\bigodot}$ the
fluctuations become comparable to its mean value when it reaches a
Schwarzschild radius of the order of $r_S\sim 10\, \textrm{nm}$
and thus its mass is still much larger than the Planck mass and
the semiclassical approximation should hold.

Therefore, according to the previous estimation the fluctuations
grow and accumulate in time, the source of this accumulation is
the non local term in the Einstein-Langevin equation, which
originates the first term on the right hand side of equation
(\ref{bh4}). This result agrees with the estimation by Bekenstein
\cite{Bekenstein} who also found this long time enhancement of the
fluctuations. It differs, however, from the estimations by Wu and
Ford \cite{Ford3} who neglected the non local term. If true, this
result seems to point to the breaking of the semiclassical
approximation well before the black hole reaches the Planck mass
and a re-examination of the evaporation process may be needed. At
the moment we have to take this result with a grain of salt, as
some of the approximations used connecting the stochastic tensor
near the horizon and far from it are not totally correct
\cite{HuRou06}. But it seems clear that more work is needed, in
particular a better approximation for the noise kernel near the
black hole horizon even within the semi-adiabatic approximation is
needed. The noise kernel must be treated as a distribution, it is
singular in the coincidence limit and for null separated points,
but it is finite if properly smeared with smooth functions
suitably integrated in time as well as in space.

\section{Summary and outlook}
\label{5}

We have reviewed the semiclassical theory of gravity as the theory
of the interaction of classical gravity with quantum matter
fields. The most important equations in this theory are the
semiclassical Einstein equations (\ref{1.1}) which describe the
back-reaction of the gravitational fluctuations in its interaction
with the quantum fields. We noticed that the theory may seriously
fail when the fluctuations on the stress tensor of the quantum
fields are significant. We have then sought an axiomatic approach
by which the semiclassical equations can be corrected in order to
take into account those fluctuations. These equations turn out to
be uniquely defined and are the Einstein-Langevin equations
(\ref{1.5}) which are linear in the metric perturbations $h_{ab}$
over the semiclassical background. These equations predict
stochastic fluctuations in the metric  perturbations induced by
the stress tensor fluctuations described by the noise kernel
(\ref{1.3}). We have also noted that the stochastic correlations
of the metric perturbations predicted by the Einstein-Langevin
equations reproduce the quantum metric correlations of the quantum
theory of gravity interacting with $N$ matter fields, in the large
$N$ expansion.

We have finally used the stochastic theory in the inflationary
cosmological context. We have computed the two-point correlation
functions of the metric fluctuations during a quasi-de Sitter
expansion induced by the stress tensor fluctuations of the
inflaton field. The results are in agreement with other approaches
to the same problem \cite{Mukhanov 92}, an approximate
Harrison-Zel'dovich spectrum is predicted. We noticed that in our
approach the quantum fields and the gravitational fields are
treated separately, and this may have some advantages to go one
step further and consider the quantum field fully, not just to
linear order. We have also considered a second application in
black hole physics. We have argued that for a large evaporating
black hole, fluctuations can accumulate over time and become
significant before reaching Planck scales. But more work is needed
to confirm this calculation and to explore its possible
consequences.

Other applications of stochastic semiclassical gravity to
semiclassical cosmology have been performed \cite{Calzetta 97a},
some including thermal fields \cite{Campos and Hu,Martin 99c}. It
has been shown that noise produced by a quantum  field on the
cosmological scale factor of an isotropic closed
Friedmann-Robertson-Walker, in the presence of a cosmological
constant, may take the scale factor from a region where it is
nearly zero to a region where it describes a de Sitter
inflationary era \cite{Calzetta 99}. Thus jumping over the barrier
by activation, this is the semiclassical analogue of the tunneling
from nothing in quantum cosmology
\cite{Vilenkin,Vil83,Vil84} and gives yet
another mechanism to produce inflation. Finally, stochastic
gravity has also been used to formulate a criteria for the
validity and stability of semiclassical gravity \cite{HuRouVer04}.
In particular it has been shown, in this context, that flat
spacetime as a background solution of semiclassical gravity is
stable.

\acknowledgements

I am very grateful to the organizers of the \textit{XXIX Spanish
Relativity Meeting, ERE-2006} for giving me the opportunity to
participate at the conference on \textit{``Einstein's legacy: from
the theoretical paradise to astrophysical observation"}, and for
their kind and generous hospitality. I am also grateful to Rosario
Mart\'{\i}n and Albert Roura for their essential contribution to
the work described here, and to Daniel Arteaga, Esteban Calzetta,
Antonio Campos and Bei-Lok Hu for many discussions during the last
years on this topic. I also thank Albert Roura for a critical
reading of the manuscript and many useful suggestions. This work
has been partially supported by the Research projects MEC
FPA-2004-04582 and DURSI 2005SGR00082.

\end{document}